\documentclass{ws-ijmpcs}
\def\bfg #1{{\mbox{\boldmath $#1$}}}
\begin{document}

\markboth{Yu.N.~Uzikov \& A.A.~Temerbayev }
{Test of time-reversal symmetry}

%
\catchline{}{}{}{}{}
%

\title{Test of time-reversal symmetry in the proton-deuteron scattering
}

\author{Yu.N.~Uzikov}

\address{Joint Institute for Nuclear Researches, Joliott-Curie 6,
Dubna, 141980, Russia\\
%
Physics Department, Moscow State University, 
Moscow, 119991 Russia\\
uzikov@jinr.ru}

\author{A.A. Temerbayev}

\address{L.N. Gumilyov Euroasian National University, K.Munaitpasov 5, Astana
010000, Kazakhstan\\
adastra.77@mail.ru}

\maketitle

\begin{history}
\received{Day Month Year}
\revised{Day Month Year}
\published{Day Month Year}
\end{history}

\begin{abstract}

 The integrated
 cross section $\widetilde\sigma$  for a special type
 of  double polarized  proton-deuteron scattering
 constitutes
 a null test for
 time-invariance violating but P-parity conserving effects.
 Using  Glauber theory for the $pd$ elastic scattering and 
different types of  
 phenomenological T-odd P-even NN-interactions we show that the contribution
 of the lowest mass  meson exchange, i.e. the $\rho$-meson,
 to the null-test signal  $\widetilde\sigma$ vanishes.
 Variation of  the cross section
 $\widetilde\sigma$  due to  strong hadronic  and  Coulomb interaction is
 studied and its energy dependence  
 is calculated in the GeV region.


\end{abstract}

\ccode{PACS numbers: 24.80.+y, 25.10.+s, 11.30.Er, 13.75.Cs}

\section{Introduction}

 Time-invariance-violating  (T-odd) P-parity conserving
 (P-even) (TVPC) 
  interactions  do not arise on the fundamental level
 within the standard model.
 This type of interaction
 can be generated by radiative corrections to the T-odd P-odd interaction
 discovered in physics of kaons   and B-mesons.
 However in such a case  its intensity  is too small to   be  observed in experiments
 at present~\cite{Khripl91}. Thus,   observation of  TVPC effects 
 would be considered as indication to physics beyond the
 standard model.

As was shown in Ref.~\refcite{conzett}, the total polarized cross section
 ${\widetilde \sigma}$ 
 of the  proton-deuteron scattering with vector polarization
 of the proton $p_y^p$ and tensor polarization
 of the deuteron $P_{xz}$ 
constitutes a  null-test of TVPC   effects.
The dedicated experiment is planned at COSY~\cite{TRIC} at proton beam energy 135 
MeV
 which is
 motivated by  theoretical estimations of the TVPC effects~\cite{beyer}.
 Analysis  of the TVPC null-test signal~\cite{beyer} 
was done within the nonmesonic deuteron breakup channel
 $pd\to ppn$ estimated in the impulse (single scattering) approximation.
 Here we use the spin-dependent  formalism~\cite{PK} of the Glauber
 theory to calculate the cross section $\widetilde{\sigma}$. 
The formalism includes  full spin dependence of elementary  pN-amplitudes and S- and D-components
   of the deuteron wave function. 
This formalism
 was used~\cite{TUZyaf}     to calculate non-polarized
   differential  cross section and spin observables  of the elastic $pd$ scattering 
   and total polarized $pd$ cross sections at energy of the TRIC experiment 135 MeV.
    We  further   develop    the formalism 
  to account for the Coulomb 
 and TVPC interactions.

\section{Elements of formalism}

The total cross section of the $pd$ scattering has the 
form~\cite{TUZyaf}
 \begin{equation}
\label{totalspin}
{ 
\sigma_{tot}= {\sigma_0+\sigma_1{ {\bf p}^{ p}\cdot {\bf p}^d}+
 \sigma_2 {({\bf p}^{ p}\cdot {\hat {\bf k}}) ({\bf p}^d\cdot {\hat {\bf k}})}+
\sigma_3 { P_{zz}}} +{\widetilde \sigma} {p_y^p P_{xz}^d},
}
\end{equation}
where ${\bf p}^p$  (${\bf p}^d$) is the  vector polarization of the  initial proton (deuteron) and 
$P_{zz}$ and $P_{xz}$ are the tensor polarizations of the deuteron. The OZ axis is directed along
 the proton beam momentum $\hat {\bf k}$, OY$\uparrow\uparrow {\bf p}^p$, 
OX $\uparrow\uparrow [{\bf p}^p\times {\hat{\bf k}}]$. 
%
%
 In Eq. (\ref{totalspin}) the terms $\sigma_i$ with $i=0,1,2,3$ are non-zero only 
 for T-even P-even interaction and
 the last term ${\widetilde \sigma}$ 
constitutes  a null-test signal of T-invariance
violation with P-parity conservation.
Hadronic amplitudes of $pN$ scattering are taken as \cite{PK} 
\begin{eqnarray}
\label{pnamp}
M_N({\bf p}, {\bf q};\bfg \sigma, {\bfg \sigma}_N)=
 A_N+C_N\bfg \sigma \hat{\bf  n} +C_N^\prime\bfg \sigma_N \hat{\bf  n }+
B_N(\bfg \sigma \hat {\bf k}) (\bfg \sigma_N \hat {\bf k})+\\ \nonumber
+ (G_N+H_N)(\bfg \sigma \hat {\bf q}) (\bfg \sigma_N \hat {\bf q})
+(G_N-H_N)(\bfg \sigma \hat {\bf n}) (\bfg \sigma_N \hat {\bf n}),
\end{eqnarray}
where  orts ${\hat {\bf q}}$, ${\hat {\bf k}}$ and ${\hat {\bf n}}$
 are defined as unit vectors along the vectors  ${ {\bf q}}={\bf p}-{\bf p}'$,
${ {\bf k}}={\bf p}+{\bf p}'$
and ${ {\bf n}}=[ {\bf k}\times {\bf q}]$,
respectively; ${\bf p}$ (${\bf p}^\prime$) is the initial (final) proton momentum. 
%
 We  consider  the  following terms of the TVPC NN interaction which were under
 discussion in  Ref.~\refcite{beyer}: 
\begin{eqnarray}
\label{TVNN}
t_{pN}={h_N[({\bfg \sigma} \cdot {\bf k})({\bfg \sigma}_N \cdot {\bf q})+
({\bfg \sigma}_N \cdot {\bf k})({\bfg \sigma} \cdot {\bf q})-
\frac{2}{3}({\bfg \sigma}_N \cdot{\bfg \sigma})
({\bf k}\cdot {\bf q}) ]}/m_p^2
+ \\ \nonumber
+g_N [{\bfg \sigma} \times {\bfg \sigma}_N]\cdot [{\bf q }\times{\bf k}]/m_p^2
+{g^\prime_N ({\bfg \sigma} - {\bfg \sigma}_N)\cdot i\,[{\bf q}\times {\bf k}]
[{\bfg \tau} \times{\bfg \tau}_N]_z}/m_p^2.
\end{eqnarray}
Here ${\bfg \sigma}$ 
(${\bfg \sigma}_N$)
 is the Pauli matrix acting on the spin state of the proton
(nucleon $N=p,n$), 
${\bfg \tau}$  
(${\bfg \tau}_N$) 
 is the corresponding matrix acting on the isospin state; $m_p$ is the proton mass.
 In the framework  of the   phenomenological meson exchange interaction
 the term $g'$ corresponds to $\rho$-meson exchange,
and $h$-term provides the axial meson $h_1$ exchange with 
$h_N=-i2G_{h}{\bar G}_{h} F(q^2)(m_h^2+{\bf q}^2)^{-1}$,
where $G_{h}$ (${\bar G}_{h}$) is the ordinary (TVPC) $hNN$ coupling constant,
$m_h$ is the h-meson mass and $F(q^2)$ is the $hNN$ vertex
form factor \cite{beyer}.

Using the generalized optical theorem  we find
${\widetilde\sigma}=-4\sqrt{\pi}Im\frac{2}{3}{\widetilde g}$.
For single scattering mechanism the  amplitude $\widetilde g$ 
vanishes within the Glauber theory. 
For double scattering mechanism with  $pN$-amplitudes (\ref{pnamp}) and (\ref{TVNN}) 
 within the S-wave approximation for the deuteron 
wave function we find 
 \begin{eqnarray}
\label{g5} 
{\widetilde g}=\frac{i}{4{\pi}m_p}
\int_0^\infty dq q^2S_0^{(0)}(q)[C^\prime_n(q)(g_p-h_p)+C^\prime_p(q)(g_n-h_n)],
 \end{eqnarray}
where $S_0^{(0)}$ is the S-wave elastic form factor of the deuteron
in notations of Ref. \refcite{PK}.

\section{Numerical results and discussion}
 The results of our calculations~\cite{TUZyaf} for non-polarized differential cross section,
 vector $A_y$ and tensor $A_{ij}$
 analyzing powers, spin correlations parameters $C_{i,j}$, $C_{ij,k}$ 
are in reasonable agreement with the available
 experimental data
 at 135 MeV and 250 MeV  in forward hemisphere.
 We found also that account for  Coulomb effects 
 improves agreement with the data 
 at these energies at small  angles  $\theta_{cm}\le 20^\circ - 30^\circ$.
 The obtained results~\cite{TUZyaf} allow one  to conclude
 that the Glauber  theory is quite suitable
 for study of the null-test signal for TVPC effects in the $pd$-scattering  because
  this signal is not
imitated
 by the strong  background caused by T-even P-even
 interactions.     
 Our approach 
 is more general than  the approximation used in 
 previous study~\cite{beyer}.

 The $g^\prime$-term gives zero contribution to ${\widetilde g}$ in Eq. (\ref{g5}).
  Account for the  D-wave does not change this result.
 Therefore,
the $\rho$-meson exchange in general 
 allowed as the lowest mass meson
 in the TVPC $NN$ interaction~\cite{simonius97}
 and  expected to dominate
 the TVPC  $NN$ interaction,
 does not contribute to
  the null-test signal ${\widetilde\sigma}$.
 Contribution of other more heavy mesons is expected to be less important due to  $NN$
 repulsive core  at short distances between nucleons.
 A microscopic T-violating optical
 potential for nucleon-nucleus interaction
 was derived in Ref. \refcite{Engel94} starting from the T-violating $\rho$-meson interaction
  between nucleons and
 the corresponding coupling constant of the
 $\rho$-meson to the nucleon $\bar g_\rho$ is widely  used \cite{simonius97,huffman}
 as a measure of intensity of the TVPC effects.
 One can see, however, for the
 $pd$ scattering
 this parameter can not be applied directly as a scale of the TVPC interactions.

\begin{figure}[htb]
\vspace{0.5cm}
\centerline{\includegraphics[width=0.75\textwidth,angle=0]{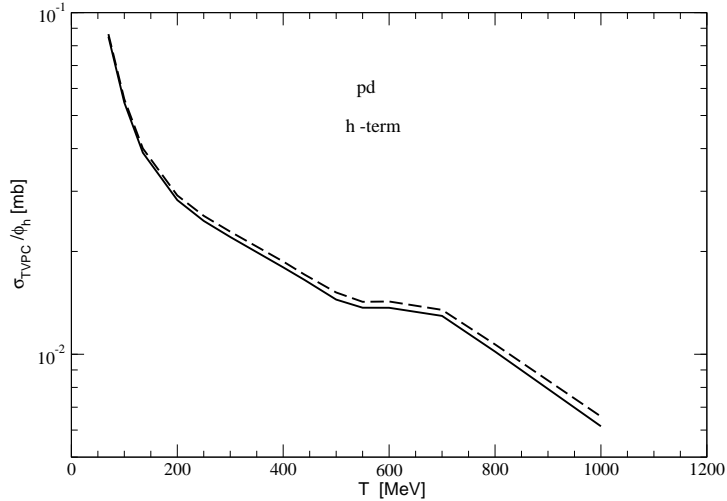}}
\caption{The calculated energy dependence of the TVPC cross section $|\widetilde\sigma|$ 
for the $h$-term in units of the constant $\phi_h=\bar G_h/G_h$ with the
  Coulomb interaction included (dashed line) and excluded (full).
}
\label{uzikov-fig1}
\end{figure}

Now we consider 
a role  of  Coulomb effects
 in the cross section $\widetilde \sigma$.
%
The Coulomb $pp$ scattering being T-even P-even interaction
 cannot generate the TVPC amplitude $\widetilde g$ within 
the single scattering mechanism, therefore its contribution 
to $\widetilde g$ is zero in this approximation.
%
To account for 
 the Coulomb interaction within the double scattering mechanism
one should add the
$pp$-scattering amplitude $f_{pp}^C$ 
to the pure hadronic 
 T-even P-even $pp$ amplitude $M_p$ in Eq.(\ref{pnamp}).
The amplitude $f_{pp}^C$ does not depend on spins and
therefore is  added only
to the spin-independent  term $A_p$.
However, the $A_p$ term
 is excluded from 
 the TVPC amplitude (\ref{g5}) due to its specific spin
 structure.
The $C_p^\prime$ is the only factor in Eq.(\ref{g5}) 
 which contains  Coulomb effects via the $A_p$ because of the relation~\cite{PK,sorensen} 
 $C_p^\prime=C_p+i({q}/{2m_p})A_p$.
 When substituting
 this amplitude 
into Eq. (\ref{g5}) and making integration over 
$q$,
 we find  that
 the singularity of the  Coulomb amplitude $f_{pp}^C(\theta_{pp})$
 at $\theta_{pp}=0$ does
 not lead to divergence of the  null-test amplitude ${\widetilde g}$.
 Energy dependence of $\widetilde\sigma$ calculated for the $h$ term
 with $G_h^2/4\pi=1.56$, $m_h=1.17$ GeV in units
 of the unknown $hNN$  coupling constant  
 $\phi_h=\bar G_h/G_h$
  is shown in Fig.\ref{uzikov-fig1}. The  $C_p$ and $A_p$ amplitudes are taken
 from Ref.\refcite{SAID}. 
 The $g$-term leads to a very similar energy dependence.
 One can see that energies $\sim 100$ MeV are more preferable than $\sim 1$ GeV
 to search for the TVPC effects.

\section{Summary}

  The Glauber theory  provides a theoretical basis for estimation
  of the TVPC effects in the double polarized $pd$ scattering
  at the energy of the   experiment~\cite{TRIC} planned at COSY.
  Using this theory we find that the $\rho$-meson exchange between nucleons
 is vanishing in the TVPC signal $\widetilde\sigma$ in  the $pd$ scattering whereas the $h$
 and $g$ terms give non-zero contributions.
  We show also that the Coulomb interaction makes a minor influence on
  this observable.




\end{document}